# TS Cache: A Fast Cache With Timing-Speculation Mechanism Under Low Supply Voltages

Shan Shen, Tianxiang Shao, Xiaojing Shang, Yichen Guo, Ming Ling, *Member, IEEE*, Jun Yang, *Member, IEEE*, and Longxing Shi, *Senior Member, IEEE*

***Abstract*— To mitigate the ever-worsening "power wall" problem, more and more applications need to expand their working voltage to the wide-voltage range including the near-threshold region. However, the read delay distribution of the static random access memory (SRAM) cells under the near-threshold voltage shows a more serious long-tail characteristic than that under the nominal voltage due to the process fluctuation. Such degradation of SRAM delay makes the SRAM-based cache a performance bottleneck of systems as well. To avoid unreliable data reading, circuit-level studies use larger/more transistors in a bitcell by sacrificing chip area and the static power of cache arrays. Architectural studies propose the auxiliary error correction or block disabling/remapping methods in fault-tolerant caches, which worsen both the hit latency and energy efficiency due to the complex accessing logic. This article proposes a timing-speculation (TS) cache to boost the cache frequency and improve energy efficiency under low supply voltages. In the TS cache, the voltage differences of bitlines (BLs) are continuously evaluated twice by a sense amplifier (SA), and the access timing error can be detected much earlier than that in prior methods. According to the measurement results from the fabricated chips, the TS L1 cache aggressively increases its frequency to 1.62× and 1.92× compared with the conventional scheme at 0.5- and 0.6-V supply voltages, respectively.**

## I. Introduction

IN RECENT years, energy efficiency has become more important for the system on chip (SoC) as the demand for Internet of Things (IoT) and other mobile devices increases in the market. Scaling down the supply voltage is one of the most commonly used methods in the low-power design, which brings the energy efficiency near to the optimal point [1]. Operating at low supply voltages, however, static random access memory (SRAM) is more prone to faults under the process variations due to its minimum-sized transistors. As a result, memories demand a bigger design margin than that of logic circuits [14]. There are two major types of failures in memory cells: 1) timing failures that increase the cell access time and 2) unstable read/write operation [2]. The later problem can be solved by using the dedicated read port in cells, such as 8T [3], [4], [24] and 10T [5]. This article focuses on the former that dramatically degrades the read performance of SRAM under the low-voltage region. A potential timing failure during both reads and writes is essentially caused by the global process variation that could weaken both P and N devices by increasing their $Vth$ [14]. In an SRAM reading, discharging the bitlines (BLs) with large capacitances through those weakened memory cells becomes slower, making the small voltage difference between BL and BLB difficult to be sensed by a sense amplifier (SA). Fig. 1 shows a 10K Monte Carlo simulation of discharging time corresponding to the BL swing of 150 mV in a 28-nm 256-row SRAM array at 0° SS corner. The parameter variations include both local and global variations of transistors used by SRAM cells (such as threshold voltage, gate oxide thickness, channel length, and width) which are defined in the foundry lib, while the voltage and temperature are fixed without any variation in this simulation. At the nominal $V_{DD}$, the worst case latency is only 135 ps to develop enough voltage swing. For the 0.5-V supply voltage, by contrast, the worst case latency extends to 30 ns to read the minor weak bits safely. Meanwhile, the mean value and standard deviation of the distribution also increase to 7.4 and 2.36 ns, respectively. This long tail of the distribution at low supply voltage indicates that an extra timing margin must be applied, which significantly limits the throughput of the low-power SRAM [13].

The increase of memory latency makes the SRAM-based cache become the main performance bottleneck of systems

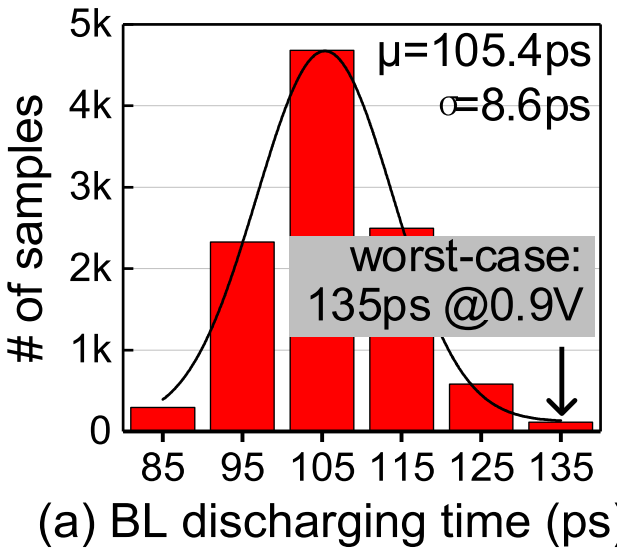

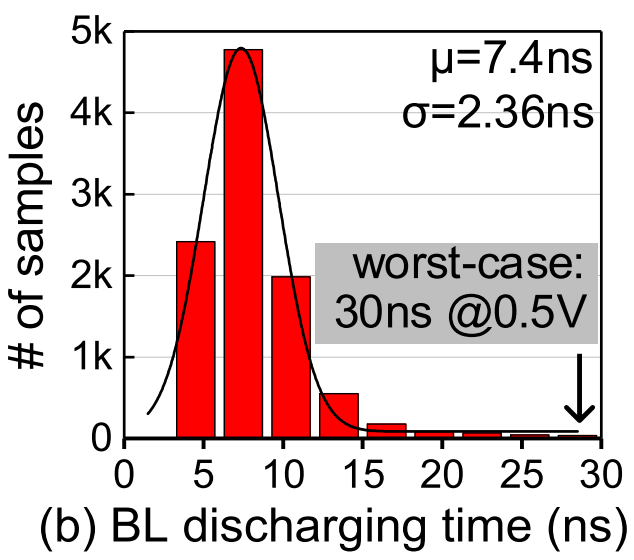

Fig. 1. 10K Monte Carlo simulation of discharging time corresponding to the BL swing of 150 mV in a 28-nm 256-row SRAM array operating at (a) 0.9- and (b) 0.5-V $V_{DD}$ 0° SS corner.

Manuscript received April 23, 2019; revised July 8, 2019; accepted August 11, 2019. Date of publication September 5, 2019; date of current version December 27, 2019. This work was supported in part by the Provincial Natural Science Foundation of Jiangsu Province under Grant BK20181141 and in part by the National Natural Science Foundation of China under Grant 61974024 and Grant 6506000195. *(Corresponding author: Ming Ling.)*

The authors are with the National ASIC System Engineering Research Center, Southeast University, Nanjing 210096, China (e-mail: shanshen@seu.edu.cn; txshao@seu.edu.cn; billionshang@seu.edu.cn; 18751959816@163.com; trio@seu.edu.cn; dragon@seu.edu.cn; lxshi@seu.edu.cn).

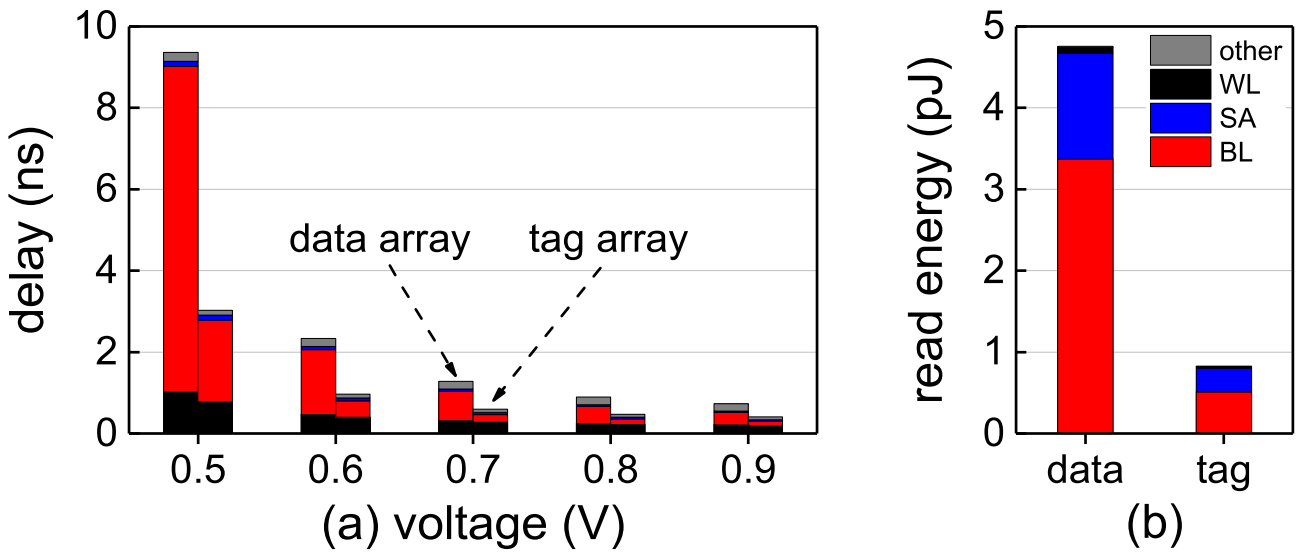

Fig. 2. (a) Read latency break down of a 28-nm 32-KB cache at different supply voltages and (b) energy break down at 0.5-V $V_{DD}$ 25° TT corner.

under low supply voltages as well. Fig. 2 shows the delay and energy breakdown of a 28-nm 32-KB L1 cache at 25° TT corner. As the supply voltage scales down, discharging the BLs in a data array accounts for 85.4% of the latency and 70.8% of the energy consumption at 0.5-V $V_{DD}$ since the data array is designed to have a larger size and longer BLs (compared with those in a tag array).

Prior work in the circuit-level improved the reliability of bitcells by using more transistors, such as 8T [4], 10T [5], and 7T/14T [6]. However, simply using larger or more transistors in bitcells [7], [10] comes at the cost of significant increases in chip area (lower density) and leakage power without any performance profit. Architectural-level solutions that tolerate faulty bits in a cache line include: 1) correcting defective bits through error-correcting codes (ECC), such as single-error correcting and double-error detecting (SECDED) and orthogonal Latin square code (OLSC) [15]; 2) disabling faulty resources (such as words, lines, and ways) [8]; 3) remapping faulty resources to create functional cache lines [9], [22]; or 4) mixing the large- and standard-sized SRAM cells in a cache [10]. To some extent, they make the tradeoff between the data error probability and the large hardware overhead or capacity loss.

Another perspective to improve the SRAM performance is the timing-speculation (TS) approaches [11]–[13] in the circuit level. Unfortunately, the method in [11] is only suitable for the SRAM with the logic dominant timing path, while the Razor SRAM [12] requires a complex roll-back mechanism in the processor pipeline to correct the error data. Moreover, they only provide a limited latency reduction due to the too-late error detections. The shared capacitors introduced in [13], on the other hand, are area hungry and need to be carefully designed to avoid failures in error detections. Furthermore, these studies target the SRAM rather than caches.

In this article, a SRAM design with a novel TS mechanism is proposed to mitigate the performance degradation of memories in the low-power scenarios. The voltage difference between BL and BLB in the SRAM array is sensed twice, called cross-sensing, far before the conservative sensing time such that the timing error can be detected much earlier than that in the work [11], [12]. Meanwhile, the cross-sensing mechanism is simpler and more area efficient than the shared capacitors in the scheme [13]. Based on such a SRAM array, we propose a TS cache that has a boosted frequency and high energy efficiency operating at near-threshold voltages. The contributions of this article are: 1) a TS mechanism that can aggressively reduce the read latency of the 6T SRAM under low voltages; 2) an L1 cache based on the proposed TS mechanism; and 3) comprehensive investigations and comparisons of the TS caches and the previous solutions.

The rest of this article is organized as follows. Section II presents related work. Section III introduces the mechanism of cross-sensing and the architecture of TS cache. Moreover, the effectiveness and robustness of the proposed solution are also discussed. Section IV presents a comparison of both cross-sensing and other TS techniques. The previous low-power fault-tolerant caches and the TS cache are also investigated. Section V shows the measurement results from the fabricated chips. Section VI outlines our conclusions.

## II. Related Work

### A. Circuit-Level Solutions

Alioto [1] presented an overview of the state of the art in ultralow-power VLSI design in a unitary framework. The design tradeoffs at various levels of abstractions are explored in this article. Regarding the low-power SRAM design, larger transistors in a memory cell average out the $V$th variability caused by nonuniformities in the channel doping and result in more robust devices with a lower probability of failure. Thus, Zhou *et al.* [7] proposed a joint optimization of cell size, number of redundant cells, and ECC strength to minimize total SRAM area while meeting target yields and $V_{DDMIN}$. Another approach is to use assist transistors in a bitcell to improve the noise margin when the supply voltage scales to the near-threshold region, such as 8T [3], [4], [24], 10T [5], and 7T/14T [6]. The 10T bitcell proposed by Calhoun and Chandrakasan [5] fundamentally solves the read static noise margin SNM problem and the write problem of 6T cells to allow subthreshold operation. However, the large-sized or 8T/10T cells significantly consume more SRAM area and static power.

### B. Architectural Solutions

From an architectural perspective, ECCs are commonly used to protect against soft errors. At low bit-error rates (BER) (e.g., only one or two fault bits in a cache line), simple ECC schemes such as parity bits or SECDED achieve good performance with small overhead. Wang *et al.* [4] observed that access-time faults occur only when a “0” bit is read on an 8T cell for a full RBL swing. Thus, they proposed the zero-counting and adaptive-latency cache (i.e., ZCAL cache) based on an 8T SRAM to detect access-time faults dynamically using a lightweight zero-bit counting error detection code. When a fault occurs, ZCAL cache extends its access time. However, considering a high bit failure rate in 6T SRAM, the simple error correction techniques cannot deal with multi-bit errors in a data chunk. Thus, a stronger ECC with larger latency, area, and energy overhead has to be applied. Chishti *et al.* [15] proposed the OLSC [15] to address both persistent and nonpersistent failures by trading off cache capacity for lower voltages. It does not rely on testing to identify and isolate the defective bits and, therefore, enables error tolerance for nonpersistent failures such as erratic bits and soft errors at low voltages. However, a large portion of the cache (25%–50%) to store the ECC check bits leads to more performance degradation (caused by more cache misses) in the low-voltage mode. A more efficient ECC method, called variable-strength error correcting codes (VS-ECCs) [19], was proposed by Alameldeen *et al.* [19]. The authors found a novel cache architecture in which only a few cache lines experience multi-bit failures at low voltages, while the vast majority of lines exhibit zero or one errors, especially for large caches. Thus, VS-ECC handles the no failure cache lines with SECDED, while using a strong 4-bit error-correcting code (4EC5ED) or a variable-length code in a small number of lines with persistent failures. However, only the cache lines with the strong ECC protection are available when the cache enters the low-power mode. Moreover, when the supply voltage changes, the characterization phase needs to be rerun to classify cache lines on the basis of the number of bit failures. For resistive memories, where errors are the result of permanent cell failures, Schechter *et al.* [20] proposed error-correcting pointers (ECP) to provide longer lifetimes by permanently encoding the locations of failed cells into a table and assigning cells to replace them.

A compromised method put forward by Khan *et al.* [10] used the heterogeneous 6T cell architecture to enable the low-voltage operation. Only clean data are stored in the nonrobust cache ways, which are protected by a simple ECC mechanism. In the case of an error, the correct data can be obtained from the lower level cache or memory. Dirty data are stored only in the robust ways constructed with larger sized memory cells, which is guaranteed by a modified replacement policy. The replacement policy, however, would incur extra cache way swapping and energy consumption. Concertina [9] allocated the faulty subblocks to the null cache subblocks, enabling the use of 100% of the last-level cache capacity. But detecting the available blocks and rearranging them in the remapping mechanisms increase the access latency and the complexity of the cache management. Hong and Kim [21] proposed a pipelined L1 cache architecture to hide the long BL discharging latency under low supply voltage by employing multi-cycle cell access and subarray-level parallel access.

### C. Timing Speculation

The concept of TS is first proposed in logic circuits to eliminate the over-design margins by *in situ* timing error detection. Alameldeen *et al.* [19] used the flip-flop and the shadow latch to double sample input data at different clock edges. The scheme is often used in a dynamic voltage scaling (DVS) system to reduce the voltage margin. Karl *et al.* [11] applied this idea to SRAM, which contains shadow SAs in addition to the main SAs. The main SA is triggered speculatively at the clock negative edge. After a while, the shadow SA resamples the BLs to confirm the result. The system detects the number of errors where the two samples are different during voltage scaling. When the number of errors exceeds the preset threshold, the supply voltage cannot be further reduced. Khayatzadeh *et al.* [12] proposed the Razor SRAM that reads memory twice with dual ports in a pipelined manner. In most cases, the read output is available after the first cycle and

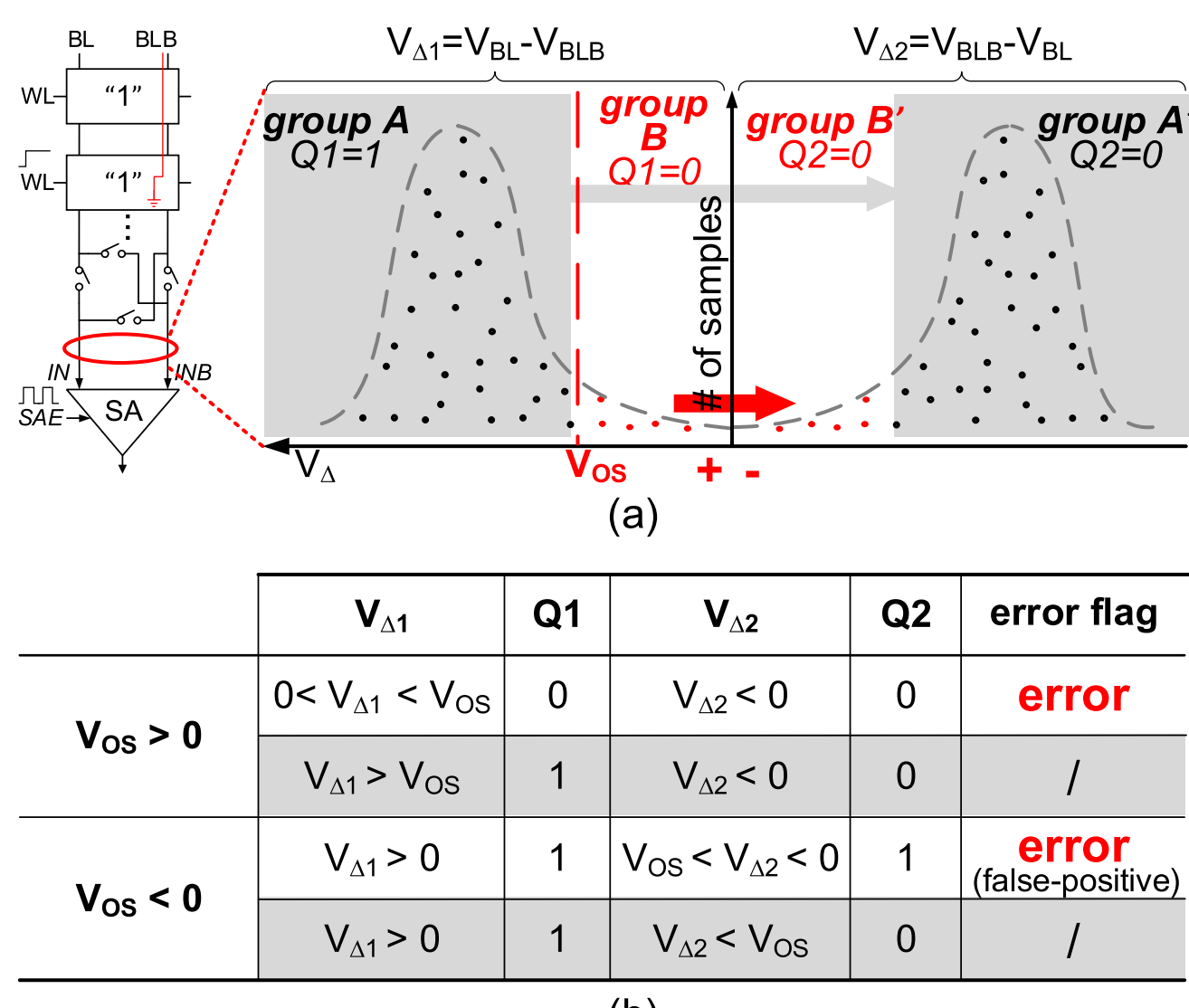

| | $V_{\Delta1}$ | Q1 | $V_{\Delta2}$ | Q2 | error flag |
|---|---|---|---|---|---|
| $V_{OS}$ > 0 | 0< $V_{\Delta1}$ < $V_{OS}$ | 0 | $V_{\Delta2}$ < 0 | 0 | error |
| | $V_{\Delta1}$ > $V_{OS}$ | 1 | $V_{\Delta2}$ < 0 | 0 | / |
| $V_{OS}$ < 0 | $V_{\Delta1}$ > 0 | 1 | $V_{OS}$ < $V_{\Delta2}$ < 0 | 1 | error (false-positive) |
| | $V_{\Delta1}$ > 0 | 1 | $V_{\Delta2}$ < $V_{OS}$ | 0 | / |

(b)

Fig. 3. (a) Mechanism of cross-sensing. Suppose that all bitcells store "1"s. (b) Truth table of error detection for different $V_{OS}$.

then confirmed by comparing with the second sample in the next cycle. For weak bits, the error flag will be triggered due to the two unequal samples. A common disadvantage of the schemes in [11] and [12] is the long-time duration between the speculative and the confirm readings. Consequently, the too-late generation of error flags limits their applications in SoC systems. For example, a complex roll-back mechanism must be implemented in the processor pipeline to correct the error data read from the Razor SRAM, which can be extravagant in a low-power processor. To solve these challenges, Yang *et al.* [13] proposed a double sensing scheme with selective BL voltage regulation (DS-SBVR), where the BL voltage is dynamically regulated by charge sharing between two sensing steps. Different from other timing speculation SRAMs, its error flag is generated much earlier. Unfortunately, the shared capacitors with large capacitances in [13] are tremendously area hungry. In addition, their capacitances must be carefully designed to avoid failures in error detection, which could possibly corrupt the data. Furthermore, all these prior studies focus on SRAM arrays rather than caches.

## III. Timing-Speculation Cache

In this section, the mechanism of cross-sensing is introduced and the overall architecture of TS cache is described. Moreover, the noise analysis is also discussed in detail.

### A. Cross-Sensing Mechanism

In the cross-sensing phase, two successive SA enable (SAE) signals are triggered and the inputs of the activated SA are switched at the second SAE. Fig. 3(a) demonstrates the mechanism of cross-sensing. Assume that the offset voltage ($V_{OS}$) of the SA is positive and the bitcells in a column store "1." The first SAE, which arrives far before the conservative sensing, activates the SA to evaluate the voltage difference between the corresponding BL and BLB ($V_{\Delta1} = V_{BL} - V_{BLB}$). Due to the process, voltage, temperature (PVT) variations, the distribution of the voltage differences (samples) consists of two groups, $A$ and $B$. The samples in *group A* are correctly read ($Q1 = 1$) because the voltage swing on BLB is large enough to be evaluated by the SA ($V_{\Delta 1} > V_{\mathrm{OS}}$). On the other hand, samples in *group B* are wrongly read as "0"s for the small BLB swing ($0 < V_{\Delta 1} < V_{\mathrm{OS}}$). After the finish of the first sensing, the SA inputs are switched and the SAE is triggered again. Thus, the second input voltage becomes negative ($V_{\Delta 2} = V_{\mathrm{BLB}} - V_{\mathrm{BL}} < 0$) and is reevaluated, which makes the samples of *group A'* and $B'$ symmetric to those of *group A* and *B* in Fig. 3(a). Since $V_{\Delta 2} < 0 < V_{\mathrm{OS}}$, the sensing outcomes $Q2$ are all "0"s. The timing error can be identified if $Q1 = Q2$ (for samples in *group B* and *B* in this case), which means that the TS cache has to extend another cycle, such that the voltage swing of BLB can be enlarged by continuously discharging, to obtain the correct sensing result. Otherwise, if $Q1 \neq Q2$, a reliable read is confirmed, the requested data can be sent out earlier than the conventional approach.

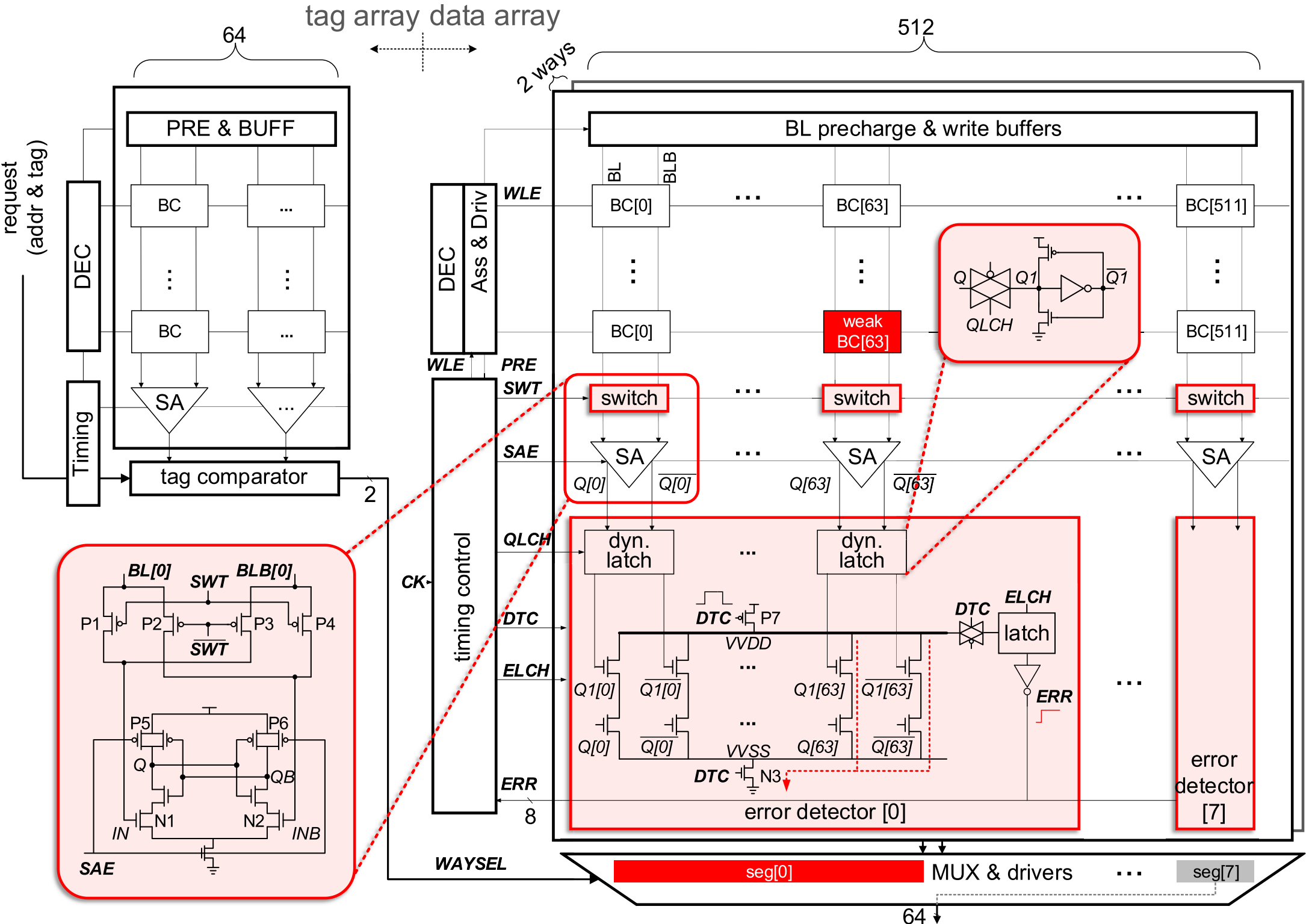

Fig. 4. Overall architecture of an instance of the TS cache with 32-KB capacity, two-way set associativity, and a 64-bit width read port.

The analogical analysis can be derived when $V_{\mathrm{OS}} < 0$, which is listed in Fig. 3(b). By using the proposed cross-sensing method, the read delay of SRAM can be aggressively improved under low supply voltages.

## B. Overall Architecture

Fig. 4 shows the overall architecture of an instance of the TS cache, which is organized as 32-KB two-way set-associativity with a 64-bit width read/write port.[1] Logically, each row of the tag array stores two 32-bit tags of the two cache ways and each row of data array stores a 64-byte cache line. In the physical layout of the TS cache, the tag and the data arrays consist of multiple subarrays, which will be shown in Section V.

In each data column, a switch comprised of four pMOS ($P1$–$P4$) is controlled by the switch (SWT) signal. When performing a normal BLs sensing, P1 and P4 are activated to connect BL and BLB to the input, IN and INB, of the SA. To swap the connections between the BLs and the SA, $P2/P3$ are turned on and $P1/P4$ are turned off. The gates of N1 and N2 in the latch-typed SA are used as the input in the case of the leakage current from BLs (SA) to the SA (BLs), which might disturb the error detections. The two pMOS transistors, $P5$ and $P6$, pull up the $Q$ and $QB$ of the SA before SAE arrives.

An error detector includes a group of dynamic latches storing the first read outcome $Q1$, and a XOR + AND gate that compares the two outcomes, $Q1$ and $Q2$, of the cross-sensing. The node VVDD that is precharged by P7 will be pulled down to the ground if any two read outcomes of a bitcell are equal (weak cell reading). At the same time, this low voltage of VVDD is latched and the error flag ERR is set. A too-large XOR + AND gate that merges the outcomes of many data columns will introduce a larger capacitance and a longer gate delay, which should be avoided in the design. Fortunately, the granularity of L1 cache data reading is set to 64 bits to match the read port width between cores and L1 caches. Accordingly, the cache line is organized by eight 64-bit wide data segments from seg[0] to seg[7], and each data segment is equipped with one individual error detector. An example when a timing error occurs in a cache line is shown in Fig. 4.

[1]To simplify our discussion in this article, we only introduce one cache configuration in this article. However, the general concepts of TS cache can be easily extended to other implementations with different cache parameters.

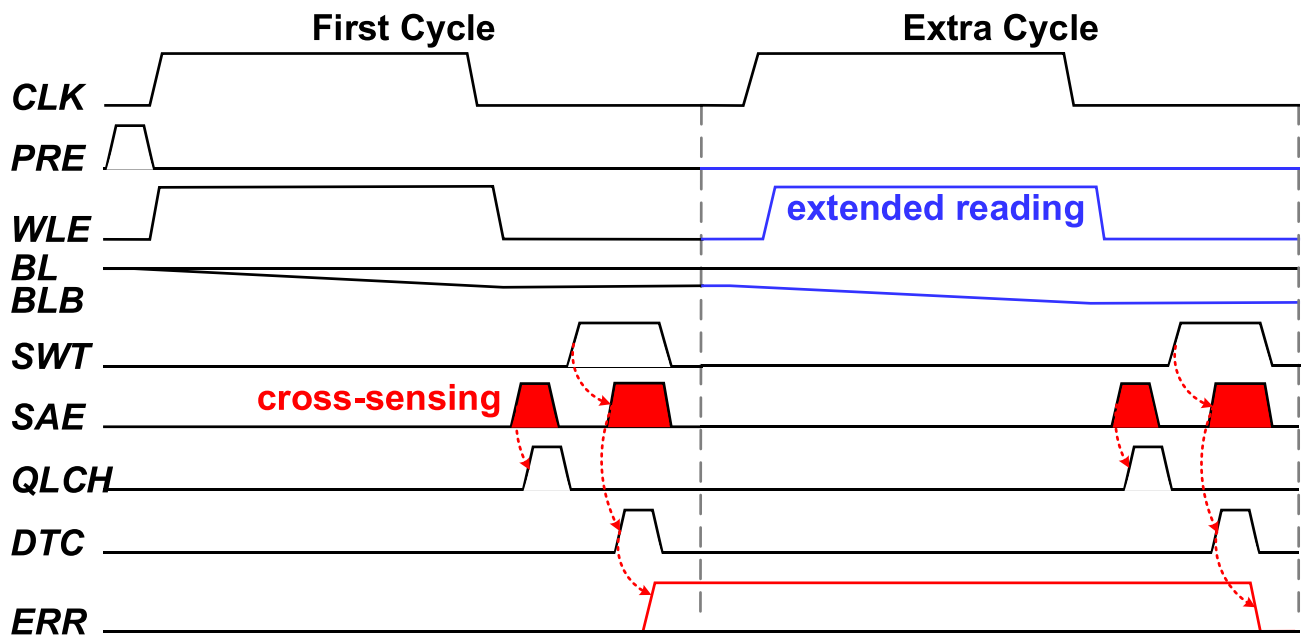

Fig. 5. Timing diagram of the TS cache.

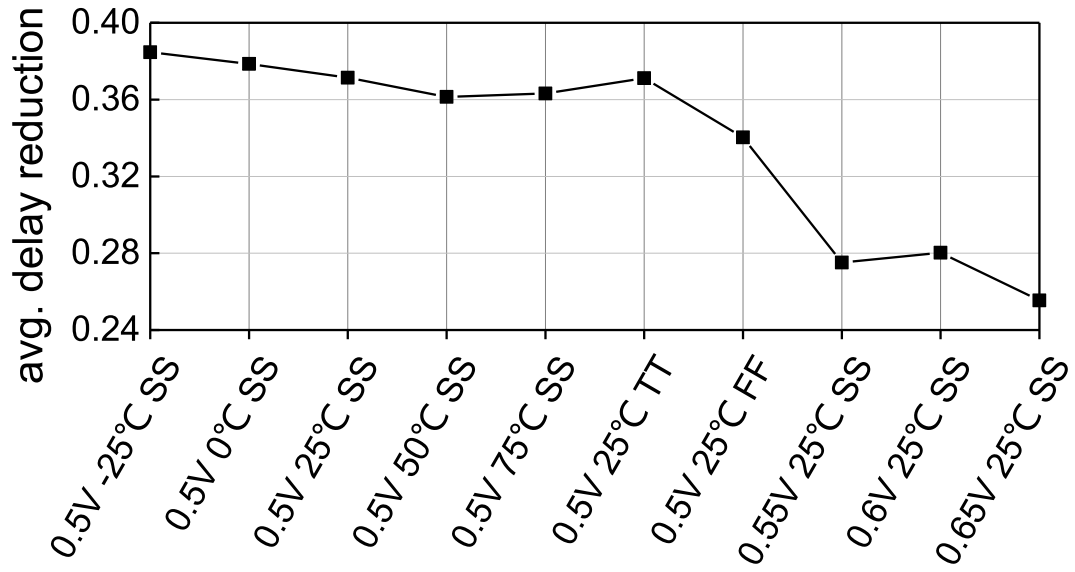

Fig. 6. Average delay reduction of cross-sensing scheme under different PVT.

The two identical sensing results from the weak cell, BC[63] in the first double-word seg[0], cause VVDD of the error detector[0] to be discharged. Then, the ERR signal of this detector is set. Since a cache often has a large width (64 bytes), there is no need to stall the whole cache line data to wait for the error-word correcting. Thus, we assume our solution is implemented in a pipelined cache where the read outcomes are temporally stored in registers. Only the subarray that contains the error word (seg[0] in this example) will perform the error correction in the next cycle, while other data segments (seg[1]–seg[7]) are transmitted to the requested core without any stalling.

Moreover, to reduce the leakage current from VVDD to VVSS, the gate length of nMOS in the XOR + AND gate is 10 nm larger than those in other modules. The dynamic latches and XOR+ AND gates used by error detectors largely reduce the area overhead compared to the static implementations.

The timing diagram is shown in Fig. 5. The SAs are activated by the first SAE signal, and the QLCH signal immediately enables the dynamic latches to store the sensing results. The SWT signal keeps high during the second SA enabling. When the second read outcome is stable, the DTC signal activates the XOR + AND gate. The ERR signal will be latched until the data are correctly read out. In most cases, the timing error can be corrected in an extra cycle by keeping the BLs discharging through the second WLE signal, shown in Fig. 5. It is possible that some extraordinary weak bits need more cycles to obtain the correct results, which may cause destructive readings in a 6T SRAM cell. However, such weaker cells can be identified by built-in self-testing (BIST) and corrected through redundancy cells or removed by block disabling. Another situation is that error data reading may occur in the nonhit cache ways. However, it is not necessary to correct these unused error data. All timing signals are generated by a configurable timing control unit with automatic PVT tracking [13], which can be flexibly configured to multiple cycles of the clock period [coming from the replica BL (RBL)].

## C. Effectiveness and Robustness Analysis

To demonstrate effectiveness and robustness of the proposed scheme, 512 Monte Carlo simulations of the TS cache with eight 256-row × 128-column sized data arrays are conducted under each combination of process corner, voltage, and temperature conditions. To limit the error correction penalty and avoid destructive reading in the 6T cell, the cross-sensing time is configured as a half of the worst case such that the error correction can be done in the next cycle. The result of the average delay reduction compared to the conventional cache without timing speculations is shown in Fig. 6. The performance improvement varies from 38% at 0.5-V –25° SS corner to 26% at 0.65-V 25° SS corner. In general, the average delay reduction increases when the $V_{DD}$ or temperature becomes lower, or under slower process corners, where the BL discharging occupies more time in a cache access (the $I_{drain}$ of BL is smaller under the worse conditions). As shown in Fig. 6, no matter how the PVT conditions change, due to the reliability and robustness of the proposed cross-sensing scheme, the timing error can always be detected and corrected in the TS cache.

Unfortunately, a false-positive situation exists in the proposed scheme, in which case the error signal is triggered while the first read outcome is actually correct. Recalling Fig. 3(b), when an SA with a negative offset voltage senses a “1,” the first output is always correct since $V_{OS} < 0 < V_{\Delta 1}$. After the SA input switching, the second read outcome can still be “1” in the condition of $V_{OS} < V_{\Delta 2} < 0$, which is caused by a small amplitude of BLB swing. Furthermore, the charge sharing between the BLs and SAs exacerbates these false positives, shown in Fig. 7(a). The equivalent capacitances of the SA input nodes IN and INB are $C_{IN}$ and $C_{INB}$, respectively. As the first SAE raises, the voltage of IN is pulled up to $V_{BL}$, while the voltage of INB equals $V_{BLB}$. After swapping the inputs of SA, the charge ($C_{BLB} \times V_{BLB} + C_{IN} \times V_{BL}$) at IN will be reshared between $C_{BLB}$ and $C_{IN}$; hence, $V_{IN}$ becomes $(C_{BLB} \times V_{BLB} + C_{IN} \times V_{BL})/(C_{BLB} + C_{IN})$. The voltage difference $V_{\Delta 2}$ can be expressed by

$$V_{\Delta 2} = \frac{C_{BLB} \times V_{BLB} + C_{IN} \times V_{BL}}{C_{BLB} + C_{IN}} - \frac{C_{BL} \times V_{BL} + C_{INB} \times V_{BLB}}{C_{BL} + C_{INB}}. \quad (1)$$

Assuming $C_{BL} = C_{BLB}$, $C_{IN} = C_{INB}$, the relation between $V_{\Delta 2}$ and $V_{\Delta 1}$ is derived as

$$V_{\Delta 2} = -\frac{C_{BL} - C_{IN}}{C_{BL} + C_{IN}} (V_{BL} - V_{BLB}) = -\frac{C_{BL} - C_{IN}}{C_{BL} + C_{IN}} V_{\Delta 1} \quad (2)$$

where the charge sharing shrinks the amplitude of $V_{\Delta 2}$ (compared to $V_{\Delta 1}$). The simulation results at 75° FF corner [Fig. 7(b)] with $C_{BL} = C_{BLB} = 50$ fF, $C_{IN} = C_{INB} = 0.5$ fF show that the absolute voltage difference is lowered

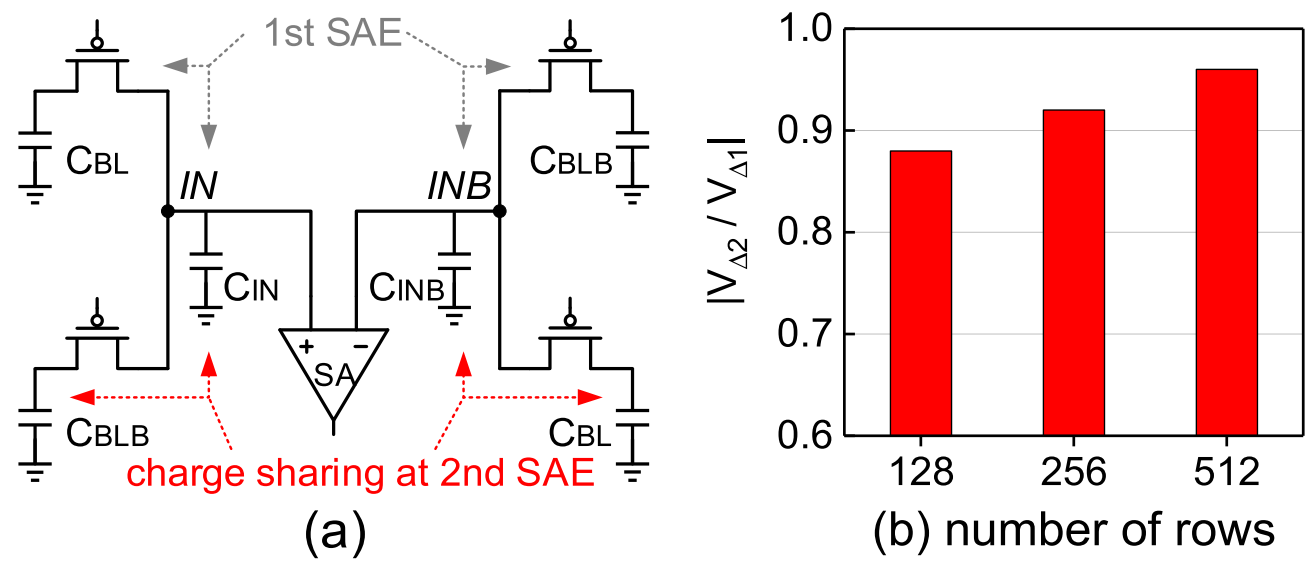

Fig. 7. (a) Charge sharing between BLs and the inputs of SA. (b) Influence of charge sharing on the second input voltage $V_{\Delta 2}$ at 75° FF corner.

TABLE I

AVERAGE DOUBLE-WORD ERROR RATE AND FALSE-POSITIVE/ NEGATIVE UNDER DIFFERENT PVT CONDITIONS

| PVT conditions | CS DER | Real DER | False-positive | False-negative |
|---|---|---|---|---|
| 0.5V -25°C SS | 1.66E-02 | 1.36E-02 | 2.99E-03 | 0 |
| 0.5V 0°C SS | 6.02E-03 | 4.92E-03 | 1.10E-03 | 0 |
| 0.5V 25°C SS | 5.78E-03 | 4.87E-03 | 9.12E-04 | 0 |
| 0.5V 50°C SS | 8.24E-03 | 6.78E-03 | 1.45E-03 | 0 |
| 0.5V 75°C SS | 5.57E-03 | 4.48E-03 | 1.09E-03 | 0 |
| 0.5V 25°C TT | 6.15E-03 | 5.16E-03 | 9.92E-04 | 0 |
| 0.5V 25°C FF | 2.61E-02 | 2.22E-02 | 3.90E-03 | 0 |
| 0.55V 25°C SS | 6.31E-02 | 5.29E-02 | 1.03E-02 | 0 |
| 0.6V 25°C SS | 2.82E-02 | 2.39E-02 | 4.25E-03 | 0 |
| 0.65V 25°C SS | 3.09E-02 | 2.57E-02 | 5.24E-03 | 0 |

by only 8% for a 256-row SRAM in the second sensing, suggesting that the increase in false positives caused by charge sharing is trivial. Table I shows the Monte Carlo simulation results of the error rates and the probability of false positives. All the probabilities of false positives in the TS cache under different work conditions are less than 1%, which means that the penalties of false-positives are trivial. This is because a false positive occurs only when all fault bits identified as timing errors in the double-word are caused by the false positives (i.e., the first sensing result of the word is actually correct but the error flag is set). Oppositely, if any bit that is detected as a failure by the cross-sensing is a real weak bit, this error detection is not a false-positive detection even though some incorrect judgments exist in the error word. Therefore, such a small fraction of false positives can be ignored in this work.

In addition, the false-negative situations, where the weak bits are wrongly recognized as the strong ones, will possibly happen in [13] when the amplitude of the regulated voltage is not sufficient. Because the false negatives destruct the data reading, the shared capacitors of DS-SBVR SRAM must be designed carefully to avoid them. Contrastively, the cross-sensing does not have this disadvantage. In the real condition, the weak bitcells, whose BL swing voltages satisfy $0 < V_{\Delta} < V_{\mathrm{OS}}$ or $V_{\mathrm{OS}} < V_{\Delta} < 0$, form a set A. In Section III-A, Fig. 3(b) shows a weak cell "1" ($V_{\Delta 1}$ is positive) is identified when $0 < V_{\Delta 1} < V_{\mathrm{OS}}$ or $V_{\mathrm{OS}} < V_{\Delta 2} < 0$. For the weak cell "0" ($V_{\Delta 1}$ is negative), error flag is set when $V_{\mathrm{OS}} < V_{\Delta 1} < 0$ or $0 < V_{\Delta 2} < V_{\mathrm{OS}}$ [not listed in Fig. 3(b)]. Combining these inequalities, the condition of cross-sensing to detect a timing failure is $|V_{\Delta}| < |V_{\mathrm{OS}}|$, which is a looser condition than that in real scenarios. According to the table, we can find that there does exist the probabilities of false positives, which are, however, very small and can be ignored in our discussion. On the other hand, the more troublesome false negatives will not happen in our scheme.

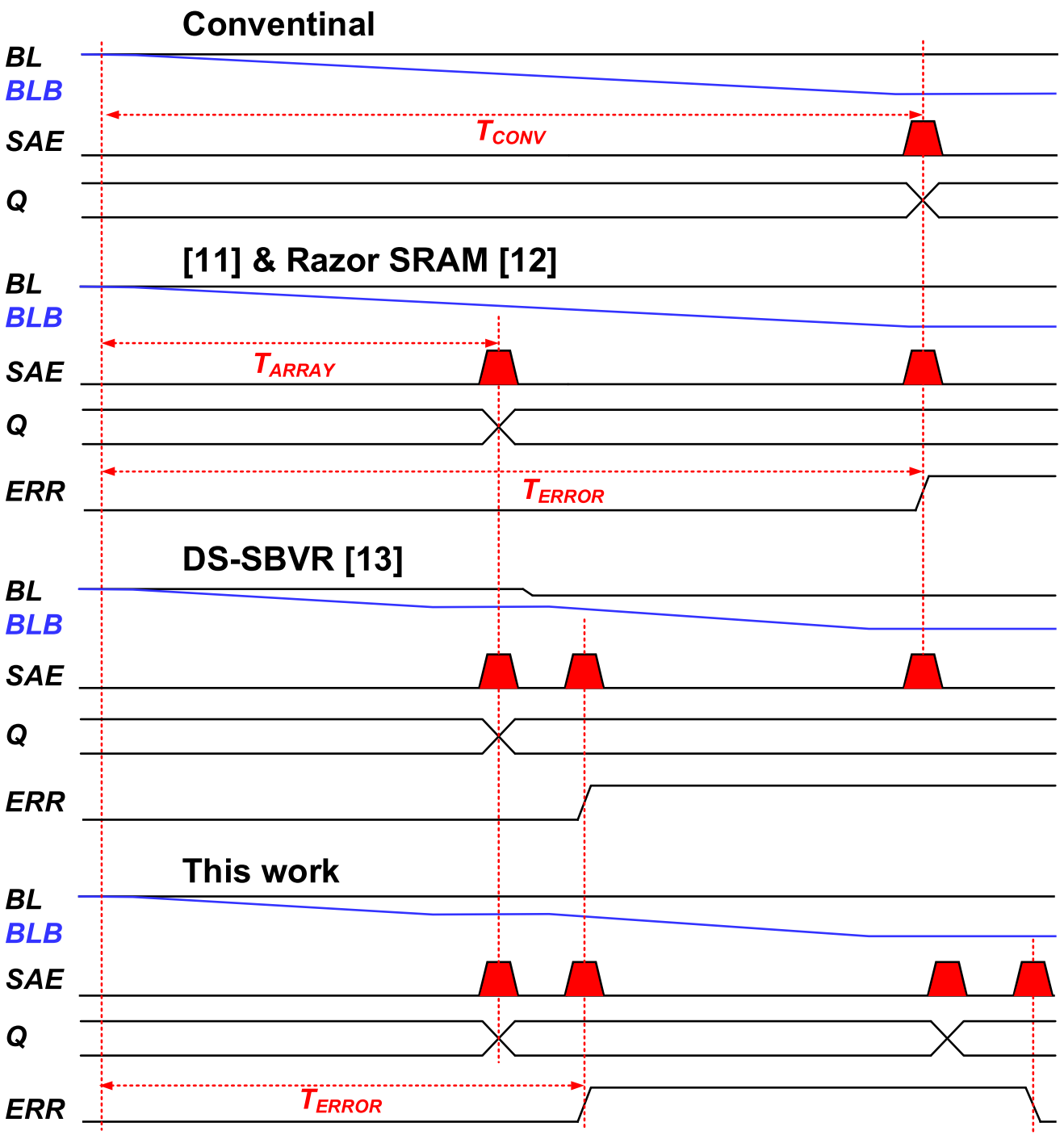

Fig. 8. Timing diagrams of different TS techniques.

## IV. COMPARISONS

In this section, comprehensive comparisons between the proposed scheme and prior approaches as well as discussion will be proceeded. To be fair, we first compare the cross-sensing scheme with other TS SRAMs. Second, the impacts on cache performance of different cache configurations will be discussed. Finally, the energy-delay product (EDP), energy, and area overhead using TS cache and other fault-tolerant caches under the low-supply voltage scenarios will be analyzed.

### A. Comparing With Other Timing-Speculation SRAMs

From a timing point of view, the speculative SRAM has two delay parameters: the $T_{\mathrm{ARRAY}}$, defined as the delay of the first speculative output, and $T_{\mathrm{ERROR}}$, defined as the delay of the final confirmation. Fig. 8 shows the comparison of these timing parameters in different speculative SRAMs. In the conventional SRAM, SA is enabled until the voltage difference between BL and BLB is sufficiently large. The delay parameter $T_{\mathrm{CONV}}$ is comprised of wordline driven, BL/BLB discharging, and SA sensing. The SRAM with the shadow SAs [11] releases the speculative outputs at the half of wordline enable time. The ideal $T_{\mathrm{ARRAY}}$ is only 50% of $T_{\mathrm{CONV}}$, while $T_{\mathrm{ERROR}}$ is equal to $T_{\mathrm{CONV}}$. However, this scheme requires that the $T_{\mathrm{ERROR}}$ must be smaller than a clock cycle to avoid the propagation of the wrong data. Thus, it is only suitable for a logic dominant path in which the logic delay occupies the most of the clock period, but not suitable for a SRAM dominant path, such as caches in processors [13]. The principle of Razor SRAM [12] is similar to [11]. Since its $T_{\text{ERROR}}$ is on a two-cycle timing path, it sends the risk data at the first cycle and detects errors in the next cycle. Therefore, it involves a roll-back mechanism when used in a processor pipeline and needs the stabilizing registers to inhibit write-backs during error detection. The ideal $T_{\text{ERROR}}$ of DS-SBVR [13] and this article is only a little larger than half of $T_{\text{CONV}}$. The maximum throughput gain is defined as the ratio of the maximum throughput to that of the conventional SRAM. The theoretical maximum throughput gains of [11] and Razor SRAM are 1.5× and 2×, and those of DS-SBVR and this article are 1.78× for the 512-row array.

TABLE II
COMPARISON WITH OTHER TS SRAMS

| | [11] | | Razor SRAM [12] | | DS-SBVR [13] | | This work | |
|---|---|---|---|---|---|---|---|---|
| Array Size | 128×32 | 512×32 | 128×32 | 512×32 | 128×32 | 512×32 | 128×32 | 512×32 |
| Sensing Scheme | Double-sensing with main and shadow SAs | | Double-sensing with dual ports in two consecutive cycles | | Double-sensing with selective bitline voltage regulation | | Cross-sensing | |
| Area Overhead | 17.6% | 4.8% | 45.1% | 50.1% | 20.8% | 7.6% | 6.4% | 1.8% |
| Energy Overhead | 52.6% | 17.9% | 56.5% | 19.0% | 34.2% | 10.1% | 12.3% | -30.6% |
| $T_{CONV}/T_{ERROR}$ | 1X | 1X | 1X | 1X | 1.57X | 1.78X | 1.6X | 1.78X |
| Max. Throughput | 1.5X | 1.5X | 2X | 2X | 1.57X | 1.78X | 1.6X | 1.78X |
| FoM in SRAM | 0.83 | 1.21 | 0.88 | 1.13 | 0.97 | 1.50 | 1.34 | 2.49 |

Moreover, the capacitances of the shared capacitors in DS-SBVR SRAM are a function of $T_{\text{ARRAY}}$, which means that the capacitors must be elaborately designed according to the timing. Oppositely, in the cross-sensing mechanism, the $T_{\text{ARRAY}}$ can be flexibly configured to achieve a different frequency boosting without any error-detection failures (discussed in Section III-C).

The compared metrics are listed in Table II. The 128-row × 32-column and 512-row × 32-column SRAM array layouts are presented using Cadence Virtuoso suit [18] in the same 28-nm process technology to demonstrate the area overhead. The bitcells in the layouts including the push-rule 6T single port (SP) and 8T dual port (DP) are provided by TSMC foundry. The baseline SRAM includes the bitcell array, SAs, and the precharge circuit without using error detection techniques. The SRAM in [11] including shadow SAs and the error detection circuits (XOR gates and MUX) consumes additional 17.6% chip area in the 128-row array. The Razor SRAM speculatively reads data through two independent ports and achieves great throughput gain at the cost of huge area overhead (45.1% for the 128-row array). The DS-SBVR SRAM has area cost of 20.8%, which is mainly consumed by the shared capacitors. Thanks to the low-cost error detector, this work achieves the best area overhead of merely 6.4% for the 128 × 32 SRAM array and 1.8% for the 512 × 32 array. The data of energy and delay are collected from simulations that are the same at 0.5-V 25° TT corner. The energy overhead refers to the ratio of the increased energy per reading operation to that of the baseline SRAM array. In an SRAM reading, energy is mainly consumed by the BL precharging, voltage sensing, and error detecting. The energy consumed by the BLs can be expressed as $1/2 \times C_{\text{BL}} \times V_{\Delta} \times V_{\text{DD}}$. As a BL connects more cells in a column, it consumes more energy due to the larger $C_{\text{BL}}$ when precharging. The cross-sensing technique proposed by this article reduces the energy consumed by BLs because of a lower voltage swing requirement (small $V_{\Delta}$). For short BLs, the energy saving of cross-sensing is smaller than the energy overhead of the additional error detection logic. On the other hand, for a larger SRAM array (e.g., 512 rows) where the BLs dominate the power consumption of the entire array, the energy saved by cross-sensing becomes larger than the energy overhead. Consequently, as given in Table II, proposed SRAM in this article reduces the reading energy by 30.6% for the 512 × 32 sized array. The energy penalties of [11]–[13], and this work are 52.6%, 56.5%, 34.2%, and 12.3% for the smaller 128-row arrays, respectively, which are mainly consumed by their error detection logic.

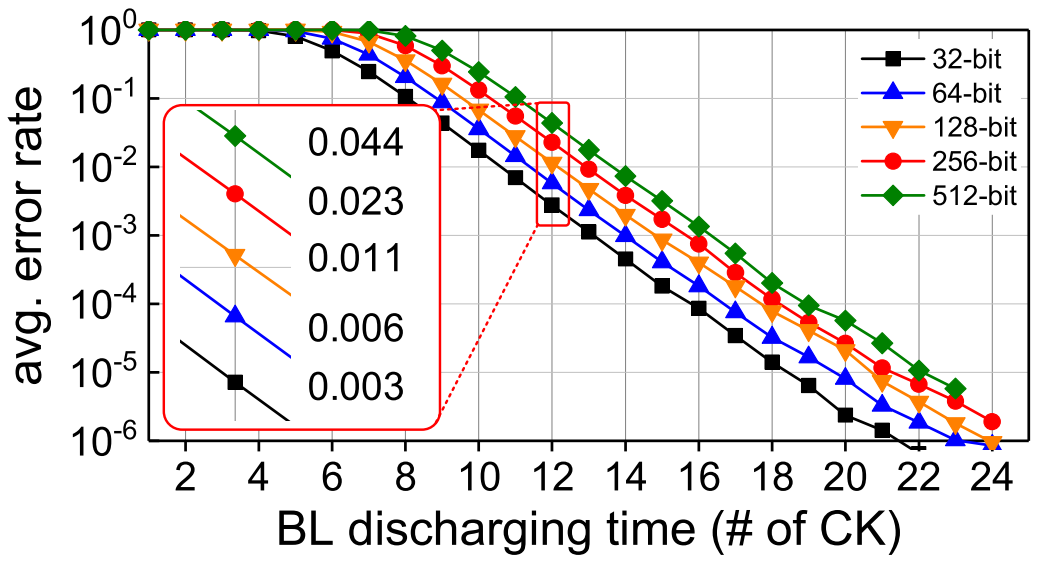

Fig. 9. Error rate (DER) detected by the cross-sensing under different word sizes in a 28-nm 256-row SRAM at 0.5-V 25° SS corner.

The figure of merit (FoM) of power, performance, area (PPA) gain is defined as the maximum throughput/(area × energy) [13]. As given in Table II, the cross-sensing scheme achieves the best FoM, 1.34 and 2.49 for the two sized arrays among all speculation SRAMs.

## B. Impacts of Different Cache Configurations

Assuming a bitcell that becomes a weak cell affected by the process variation is independent of other cells, the probability of a timing failure happened in a data word that is being read is $1 - (1 - \text{BER})^n$, where the BER is the BER in the SRAM array, and $n$ is the segment width. Fig. 9 shows the error rate under different read granularities (word size) at 0.5-V 25° SS corner averaged from 4K Monte Carlo simulations for each size. If the BL discharging time is set to 12 CKs, the word error rate varies from 0.003 for 32-bit width to 0.044 for 512-bit width. More error words introduce more frequent error correcting, leading to a nullification of the cache frequency boosting. Fortunately, the read port of L1 caches is usually narrower (64-bit in our example) than that of the L2 and L3 caches. Therefore, the penalty of error correction of the TS cache has little impact on the overall cache performance as well as the energy consumption.

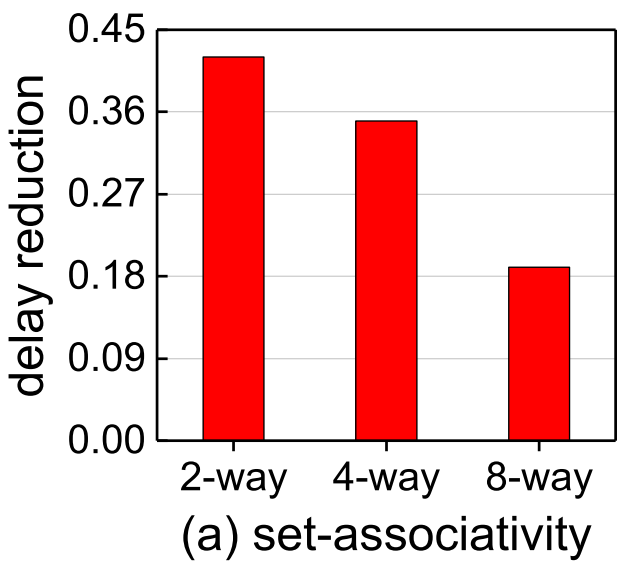

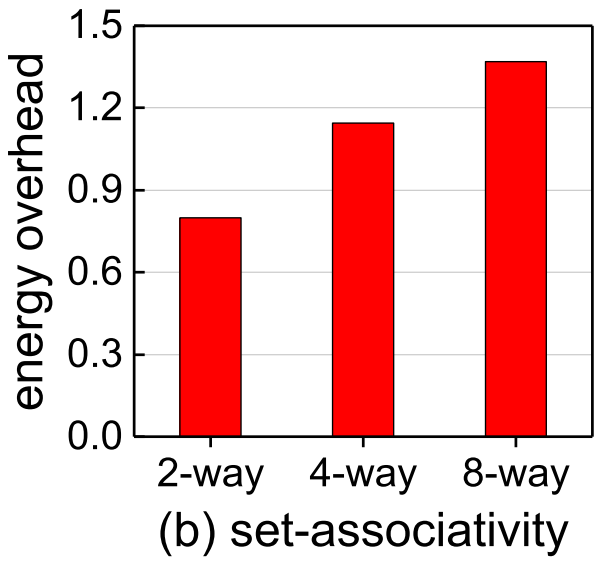

Fig. 10. (a) Read delay reduction and (b) normalized (to the baseline caches) energy overhead of TS caches with different associativities.

From Table II, it can be found that the size of the SRAM array largely affects the benefit of TS. Since all cache lines in a cache set are read simultaneously to achieve better frequency in an L1 cache design, a cache set must reside in the same row in multiple SRAM arrays according to its set index. Larger set-associativity has fewer cache sets (fewer rows) and more cache lines per set (more columns). Thus, the cache associativity actually determines the subarray size given that the cache capacity and subarray count are fixed. Three different set-associativities, 8-/4-/2-way, of a 32KB TS cache are evaluated. The corresponding subarray sizes are 64 rows × 512 columns, 128 rows × 256 columns, and 256 rows × 128 columns, respectively. Fig. 10 shows the delay reduction and the energy overhead for these three configurations. The delay reduction and the energy overhead are compared to the conventional two-/four-/eight-way set-associative caches that have $6\sigma$ correct reading probabilities without using any error detection and correction techniques (baselines). Recalling the results in Table II, a shorter BL leads to a lower BL discharging delay. Consequently, the delay reduction of the TS cache decreases as the associativity increases. Meanwhile, the larger set-associative configuration consumes more energy because more data words will be read in a cache set. Furthermore, a higher associativity also means a larger energy footprint consumed by the tag comparisons. Therefore, in a real cache design, how to arrange the SRAM subarrays in a cache is a design tradeoff. Fig. 10 also proves that the array with a square shape, e.g., 256-row × 128-column, is the most energy efficient. This array size is also adopted by the commercial caches [23] as well as other studies [24].

On the other hand, too many data subarrays increase the length of data movement in a cache. Meanwhile, more interconnections also aggressively increase the layout complexity of metal wires and decoders, which also increase the data movement delay and energy among subarrays. Since this article targets a low-power system, we use a two-way set-associative cache with eight 256 × 128 sized data arrays.

### C. Comparisons With and Other Fault-Tolerant Caches

*1) Experimental Setup:* In this work, all caches are implemented as 28-nm single banks with 32-KB capacity and two-way set-associativity[2] in the 28-nm process. The timing design of caches is according to the Monte Carlo simulations using HSPICE at 0.5-V 25° TT process corner to achieve the target yield. In the baseline version, the wordline enable time has a large margin to achieve $6\sigma$ correct reading probability without using any error detection and correction techniques. Regarding the fault-tolerant caches, the WL enable time is configured to deliver the 1 BER ($3\sigma$ correct reading probability). The energy dissipation is collected from the simulations of eight data arrays and four tag arrays. The size of each data array is 256 × 128, while the tag array size is 64 × 64.

The TS cache is compared with four other fault-tolerant caches: the mixed-cell L1 [10]; the ZCAL cache [4]; and the caches with SECDED and OLSC ECC [15]. In the mixed-cell L1 cache [10], the robust cells are designed to have 2× size after our evaluation. One of the cache ways is constructed with the larger robust bitcells while another uses the standard cells. The ZCAL cache uses 8T cells with single-ended read port and eight check bits for each 128-bit data segment. For the ECC caches, a segmented SECDED (21, 16) scheme is implemented, which can correct 1-bit error out of the 16-bit data segment with five check bits (the probability of more than 2-bit error in a segment is $P(\text{error} > 2) = 1.8\text{e}^{-9}$). The check bits of SECDED are stored together with the normal data word forming a larger SRAM array. We also evaluate a more complex ECC solution, the segment OLSC (128, 64) ECC, which reduces the probability of uncorrectable errors to $P(\text{error} > 4) = 7.1\text{e}^{-11}$ in a 64-bit data segment. However, this ECC scheme sacrifices area and power consumption since the check bits are stored in a dedicated 32-KB memory. The overhead of ECC methods refers to the results in [16].

*2) Evaluation:* The EDP is defined as the product of the single access energy and the average access latency [4], which is a lower-is-better metric. Fig. 11 shows the normalized EDP and the average read latency of different fault-tolerant schemes. The TS cache has the best EDP of 0.31 compared to the baseline. The large overhead of OLSC cache accounts for the largest EDP (0.59). The mixed-cell and TS caches improve the average read latency by 51.5% and 49.1%. For the TS scheme, the read penalty comes from the error corrections with extra cycles. The delay overhead of ECC is also very small due to the short critical path in the encode/decode logic. At a lower BER, the simple ECC solutions and TS cache have a similar performance benefit.

[2]The small associativity does hurt the performance of mixed-cell cache. However, for other low-power cache designs, the overheads are independent of the associativity of cache (they are organized at the cacheline or data segment granularity). In addition, if we hold the data array size, increasing the associativity means the cache banks must be increased in the same way. Thus, the overhead of low-power caches discussed in this article (except for mixed-cell) will change in the same way when the associativity changes.

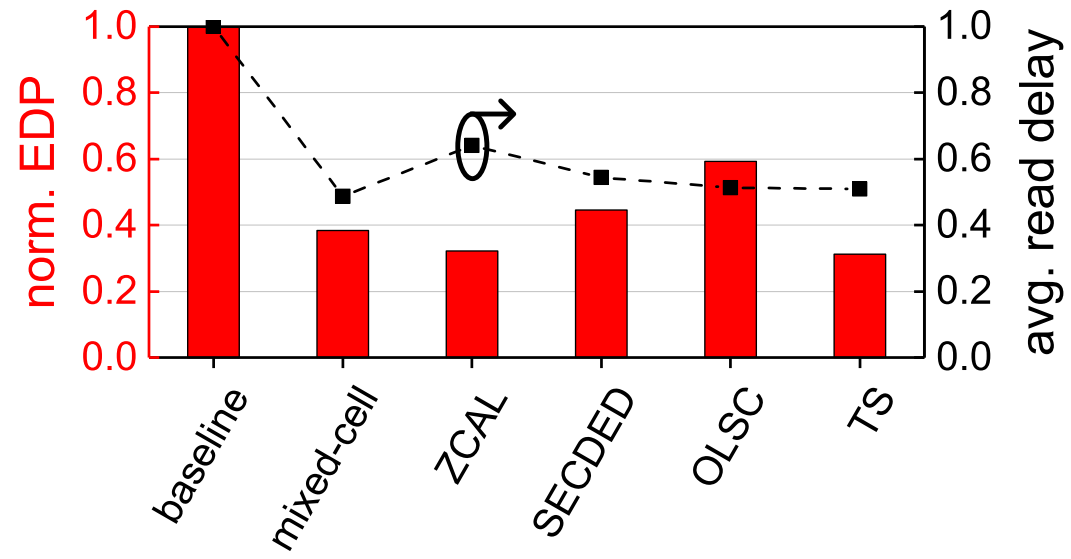

Fig. 11. EDP and the average read delay normalized to the baseline for different fault-tolerant caches.

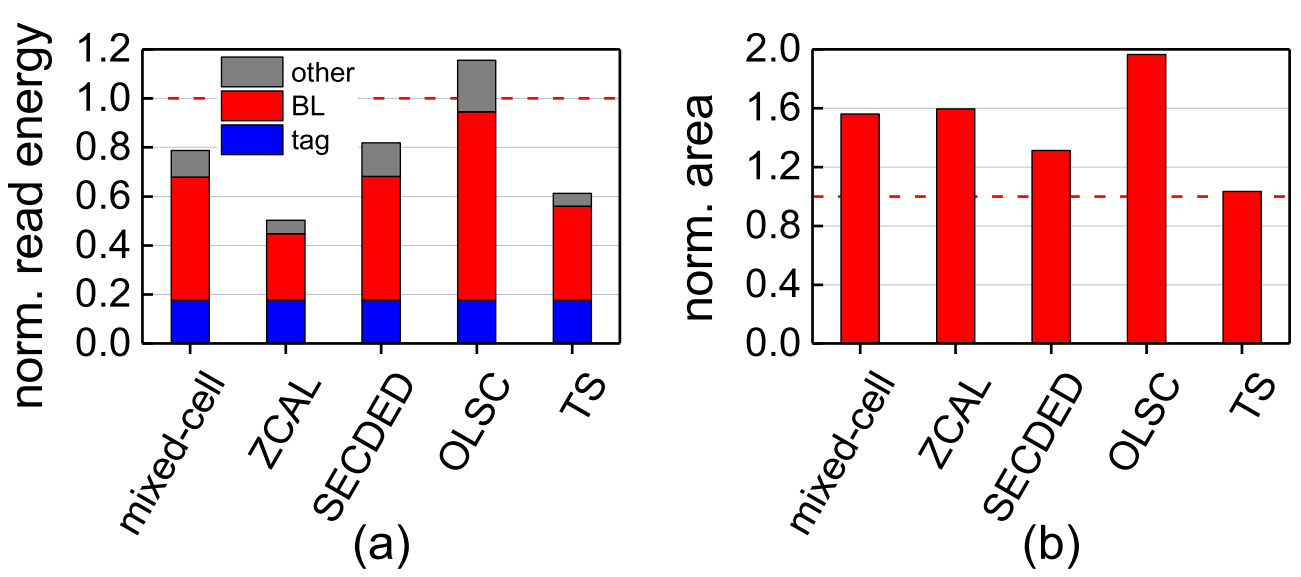

Fig. 12. (a) Energy per cache reading and (b) area for different fault-tolerant caches, normalized to the baseline version.

Fig. 12(a) shows the normalized read energy and area overhead. By using 8T bitcells, ZCAL cache [4] performs the lowest energy dissipation, only 0.5× compared to the baseline. It can be explained by the reduced frequency of reading "0" which requires a full RBL swing in ZCAL cache. Among the solutions based on the 6T SRAM, the TS cache performs the highest energy efficiency. The mixed-cell L1 and SECDED cache consume more energy due to their larger SRAM arrays. Regarding the segment OLSC (128, 64) ECC, the dedicated memory makes OLSC cache consume 1.15× energy compared to the baseline and nearly 2× compared to the TS cache. Fig. 12(b) shows the normalized chip area. As we have expected, the OLSC cache consumes 2× chip area. Meanwhile, the ZCAL cache and the mixed-cell cache also have a large area overhead. Oppositely, the TS scheme has the smallest area thanks to the limited assist hardware.

Concluding the comparisons in this section, the TS cache has the following advantages: 1) timing speculation solution provides higher energy efficiency, in which the area and energy overhead is relatively low; 2) the margin of supply voltage and timing can be aggressively harvested as long as the error bits can be read out correctly in the following extended cycles; and 3) although performance improvement of TS cache is limited in some situations, it can work correctly in various PVT conditions and a wide range of error rates (Fig. 6), where the ECC solutions cannot.

## V. MEASUREMENTS

This article describes a 32-KB single-cycle two-way set-associative TS cache prototype is fabricated with a 28-nm TSMC technology, which consists of eight data arrays with the size of 256 rows × 128 columns, a 64-bit width read/write port, and four tag arrays with the size of 64 rows × 64 columns. Fig. 13(a) shows the die micrograph and (b) depicts the testing logic of the chip. The test controller generates the chip enable signal (CEN) and the write enable signals (WEN). To mimic the cache behavior in a processor, the requested addresses and data are preprogramed in the controller. Before all cache lines being accessed sequentially, the data "0 × 55" and "0 × AA" are written into each byte of the cache by address traversal. The read outcome (Q) is sent to the comparator to count the number of error bits and error words. The WL enable time can be configured by the timing control module in the TS cache to achieve various access delays. The testing logic repeats these procedures when the timing configuration or supply voltage changes. The internal CK is generated from the RBL [13] and input to the timing control in TS cache. If any error occurs, the clock is gated to wait for the correct data. All measurement results are collected from 20 chips at room temperature (25 °C).

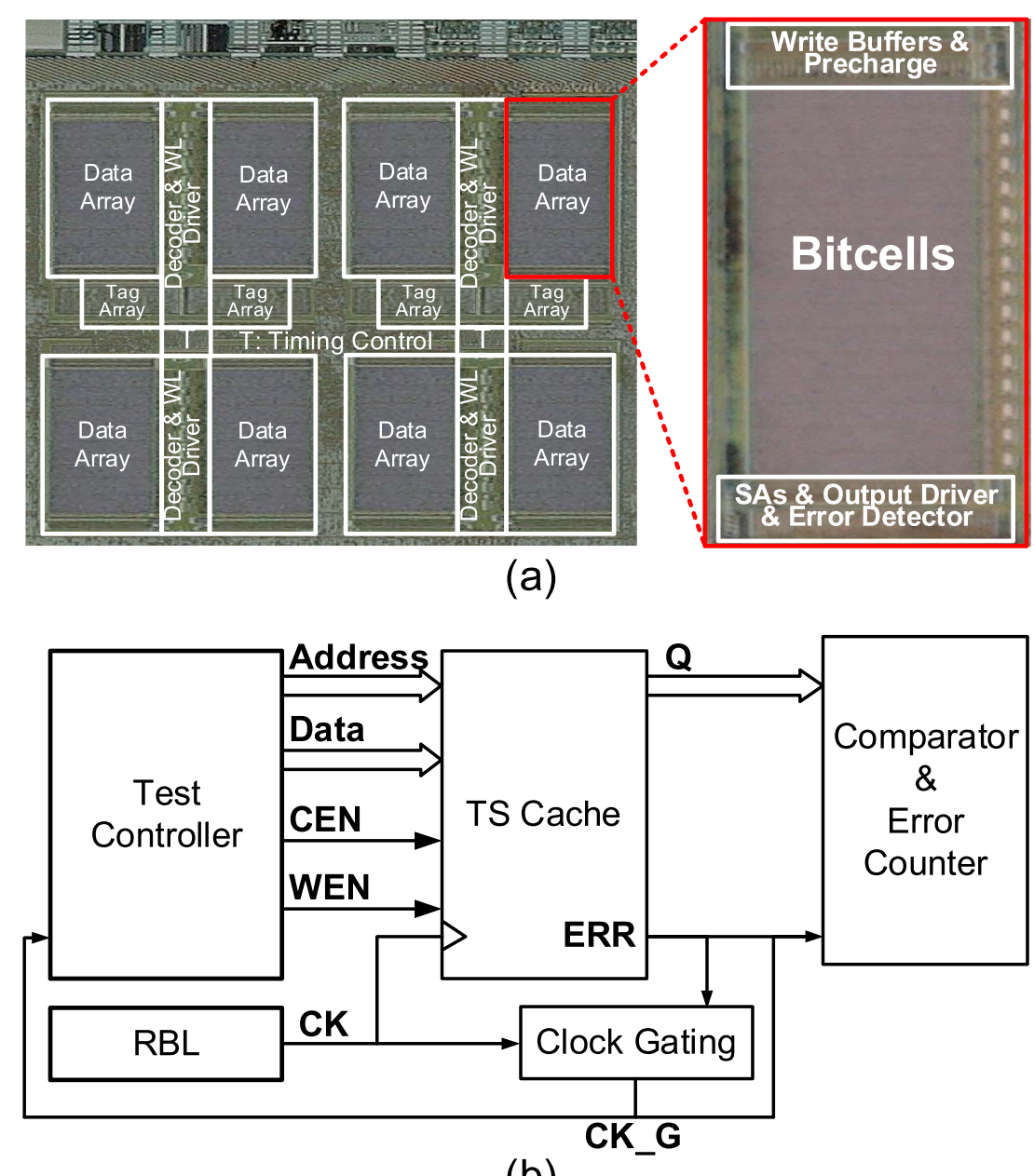

Fig. 13. (a) Die micrograph and (b) testing logic of chips.

TABLE III
CK PERIODS IN MEASUREMENTS

| VDD | Avg. (ns) | Max (ns) | Min (ns) |
|---|---|---|---|
| 0.5V | 0.687 | 0.744 | 0.658 |
| 0.6V | 0.265 | 0.279 | 0.254 |
| 0.7V | 0.167 | 0.172 | 0.161 |
| 0.8V | 0.122 | 0.125 | 0.119 |
| 0.9V | 0.099 | 0.108 | 0.096 |

Table III lists the CK periods generated by the RBL module. At 0.5 V $V_{\rm DD}$, the CK periods vary from 0.744 to 0.658 ns. Fig. 14 shows BER at different WL enable times (= the number of CKs × CK period) at different $V_{\rm DD}$. Obviously, the BER curves with longer and flatter tails as the supply voltage scales down indicate that an extremely large timing margin is indispensable to ensure the target reading yield. For example, it takes 19.23 ns (= 28 × 0.687 ns) at 0.5 V and 5.3 ns (= 20 × 0.265 ns) at 0.6 V of the WL enable time to read all cache content correctly (total 20 × 32K × 8 testing bits). Fig. 15(a) shows the average throughput gain of the TS cache. As Section IV-B illustrates, we configure the WL enable time to achieve the $10^{-3}$ BER where the best benefit point (77%) is at 0.6-V supply voltage. For 0.5-V supply voltage, the higher DER (nearly 10%) that brings more penalties to extend reading nullifies the performance benefit of frequency boosting. Compared with the baseline cache, as shown in Fig. 15(b), the frequency of TS cache is boosted by 1.6× (100 MHz) at 0.5 V and 1.9× (350 MHz) at 0.6-V $V_{DD}$ with merely 3.72% die area overhead.

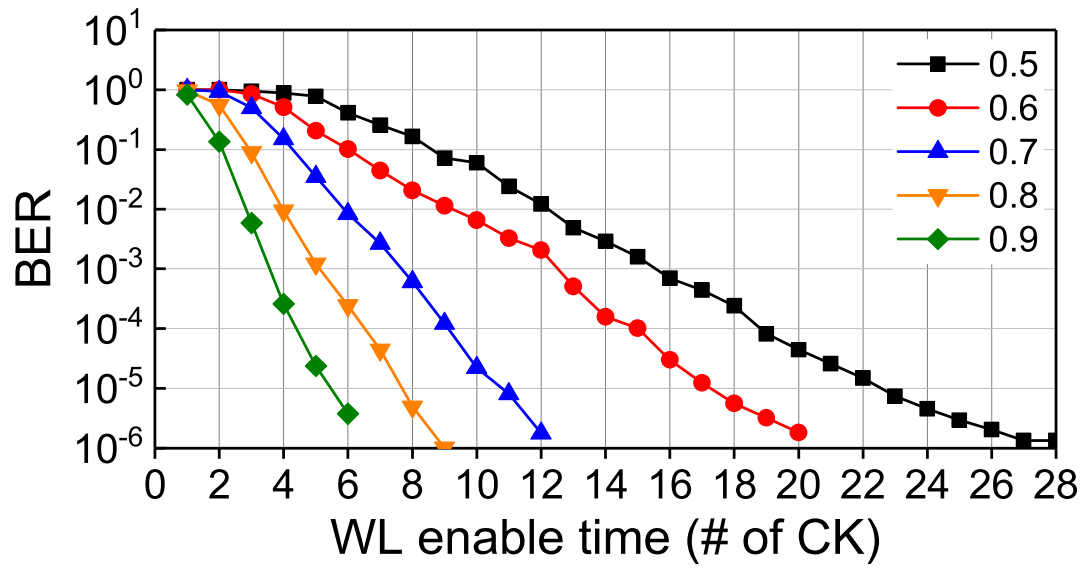

Fig. 14. BER from measurement chips at different supply voltages.

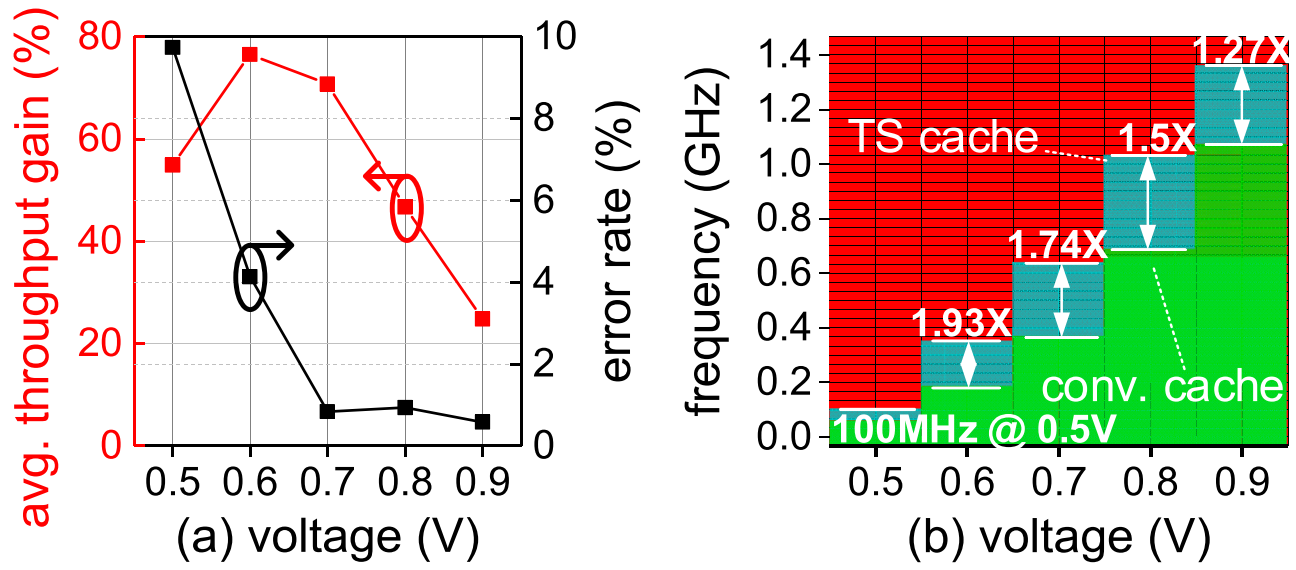

Fig. 15. (a) Throughput improvement and the DER at different $V_{DD}$. (b) Shmoo plot.

## VI. Conclusion

To address the problem of cache performance degradation under near-threshold voltage region, the TS cache is proposed in this article. By using a highly efficient TS mechanism, this article breaks through the limitation that all memory accesses must be completely correct. The erroneous reading can be quickly identified by the low-cost error detector and be corrected in an extended cycle. A 28-nm TS cache prototype is fabricated to demonstrate the effectiveness and efficiency of this scheme. According to the measurements results, the TS cache can aggressively improve the cache throughput and frequency under the low-voltage region. Beyond that, based on the standard 6T SRAM array, TS cache consumes lower chip area and energy as well. This article also conducts comprehensive comparisons with existing TS SRAMs and fault-tolerant caches including both circuit- and architecture-level solutions. All the results show that the TS cache has better energy efficiency.

In our future work, we will study on the PVT-autotracking memory based on this article in the transient PVT conditions. Combining with the DVFS techniques, which extend the application of the timing speculation cache, is also one of our future studies.

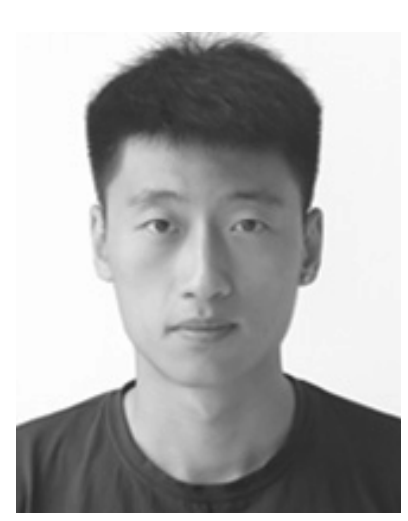

**Shan Shen** was born in 1993. He received the B.S. degree from the Microelectronics Department, Jiangnan University, Wuxi, China, in 2016. He is currently pursuing the Ph.D. degree with the School of Electronic Science and Engineering, Southeast University, Nanjing, China.

His current research interests include hardware designs in computer architecture and memory systems.

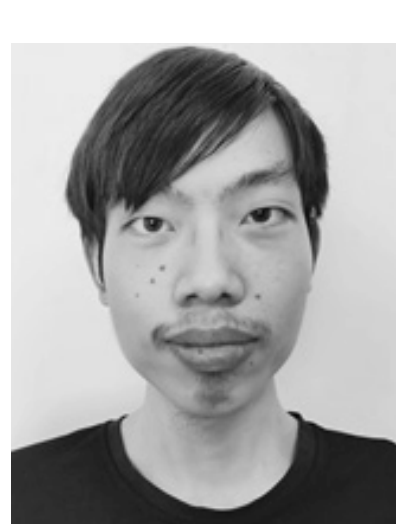

**Tianxiang Shao** was born in 1995. He received the B.S. degree from the Optics and Electronic Department, China Jiliang University, Hangzhou, China, in 2017. He is currently working toward the M.S. degree at the School of Microelectronics, Southeast University, Nanjing, China.

His current research interests include the architecture of CPU, cache, and SRAM.

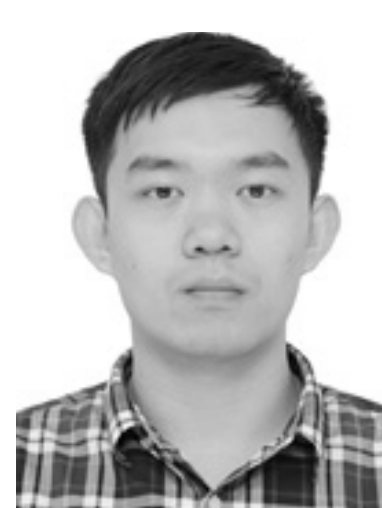

**Xiaojing Shang** was born in 1996. He received the B.S. degree from the Microelectronic Department, Jiangnan University, Wuxi, China, in 2017. He is currently working toward the M.S. degree at the School of Microelectronics, Southeast University, Nanjing, China.

His current research interests include the architecture of CPU, cache, and DRAM.

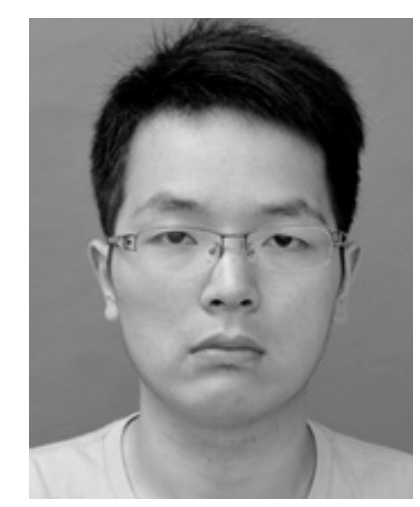

**Yichen Guo** received the B.S. degree from Anhui University, Hefei, China, in 2015. He is currently working toward the M.S. degree at the School of Electronic Science and Engineering, Southeast University, Nanjing, China.

His current research interests include low-voltage static random access memory (SRAM) circuit design.

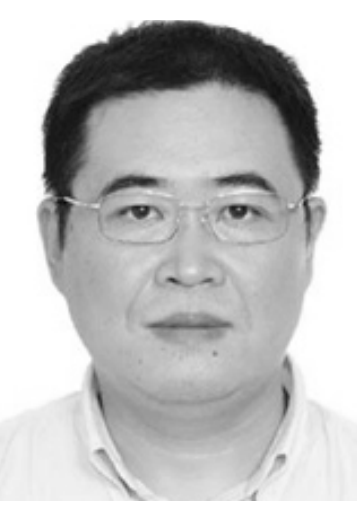

**Ming Ling** (M'19) received the B.S., M.S., and Ph.D. degrees from Southeast University, Nanjing, China, in 1994, 2001, and 2011, respectively.

His current research interests include memory subsystems of SoC, embedded software, and SoC architecture.

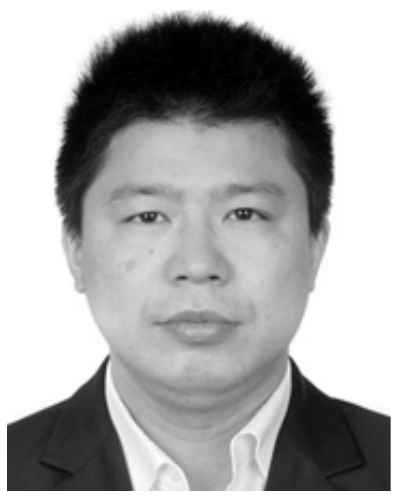

**Jun Yang** (M'15) received the B.S., M.S., and Ph.D. degrees from Southeast University, Nanjing, China, in 1999, 2001, and 2004, respectively.

He is currently a Professor at the School of Electronic Science and Engineering, Southeast University. He has coauthored more than 50 academic papers. He holds 40 patents. His current research interests include near-threshold circuit design and the Global Navigation Satellite System (GNSS) algorithm.

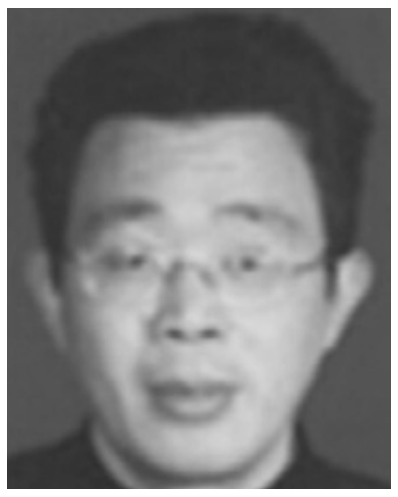

**Longxing Shi** (SM'06) received the B.S., M.S., and Ph.D. degrees from Southeast University, Nanjing, China, in 1984, 1987, and 1992, respectively.

From 1992 to 2000, he was an Associate Professor with the School of Electronic Science and Engineering, Southeast University, where he has been a Professor and the Dean of the National ASIC System Engineering Research Center, since 2001. He has authored one book and more than 130 articles. His current research interests include ultralow-power IC design.